\documentclass[
showpacs,preprintnumbers,amsmath,amssymb]{revtex4}
\usepackage{amsmath}
\usepackage{empheq}
\usepackage{graphicx}
\begin{document}

\title{INFLATIONARY UNIVERSE IN TERMS OF A VAN DER WAALS VISCOUS FLUID}

\author{I. Brevik$^{1}$\footnote{E-mail: iver.h.brevik@ntnu.no},
E. Elizalde$^{2,3,4}$\footnote{E-mail: elizalde@ieec.uab.es},
S. D. Odintsov$^{2,4,5}$\footnote{E-mail: odintsov@ieec.uab.es},
A. V. Timoshkin$^{6,7}${\footnote{E-mail: alex.timosh@rambler.ru}
}}

\medskip

\affiliation{$^{1}$Department of Energy and Process Engineering, Norwegian
University of Science and Technology, N-7491 Trondheim, Norway\\
$^2$Consejo Superior de Investigaciones Cient\'{\i}ficas, ICE/CSIC-IEEC,
Campus UAB, Carrer de Can Magrans s/n, 08193 Bellaterra (Barcelona) Spain\\
$^3$Kobayashi-Maskawa Institute, University of Nagoya, Nagoya, Japan\\
$^4$International Laboratory for Theoretical Cosmology, Tomsk State University
of Control Systems and Radioelectronics (TUSUR), 634050 Tomsk, Russia\\
$^{5}$ICREA, Passeig Luis Companys, 23, 08010 Barcelona, Spain\\
$^6$Tomsk State Pedagogical University, Kievskaja St 60, 634050 Tomsk, Russia\\
$^{7}$National Research Tomsk State University, Tomsk 634050, Russia},

  \today

\begin{abstract}
The inflationary expansion of our early-time universe is considered in
terms of the
van der Waals equation, as equation of state for the cosmic fluid, where a bulk
viscosity contribution is assumed to be present. The corresponding
gravitational equations for the energy density in a homogeneous and isotropic
Friedmann-Lema\^{\i}tre-Robertson-Walker universe are solved, and an analytic expression for
the scale factor is obtained. Attention is paid, specifically, to the role of
the viscosity term in the accelerated expansion; the values of the slow-roll
parameters, the spectral index, and the tensor-to-scalar ratio for the van der
Waals model are calculated and compared with the most recent astronomical data
from the Planck satellite. By imposing reasonable restrictions on the
parameters of the van der Waals equation, in the presence of viscosity, it is
shown to be possible for this model to comply quite precisely with the
observational data. One can therefore conclude that the inclusion of
viscosity in the theory of the inflationary epoch definitely improves the cosmological
models.

\end{abstract}

\pacs{98.80.-k, 95.36.+x}
  \maketitle
\section{Introduction}

Recent observational data from the Planck survey \cite{ade13,ade13a} confirm
the existence of an early-time accelerated period, called inflation
\cite{linde08,gorbunov11}.  In the inflationary epoch the universe expands very
rapidly, because both the total energy and the scale factor increase
exponentially \cite{brandenberger11} (a review of the universe
acceleration in modified  gravity
is given in \cite{nojiri07a,capo}). The inflationary
phenomenon can be described by using a
perfect fluid model, by $F(R)$ gravity or by a combination of both
\cite{bamba14}. The  accelerated
expansion of the late-time universe can be expressed in terms of a dark fluid,
satisfying an either homogeneous, or inhomogeneous, equation of state. One
observes that the stage of accelerated expansion is a common property of the
inflationary and late-time stages of our universe. It becomes therefore
possible to formulate the theory of the inflationary epoch similarly to the one
that holds valid for the late-time evolution of the cosmos. The inflationary epoch,
described in terms of a perfect fluid satisfying an inhomogeneous equation of
state \cite{nojiri05} can be considered as a modified theory of gravity
\cite{nojiri07a}. Note that viscous fluids represent a particular class
of the fluids with an inhomogeneous equation of state  \cite{nojiri05}.

Investigations about viscous cosmology started some time ago
but their application to the accelerating universe are rather recent.
We may mention some relevant papers, as
\cite{weinberg71,padmanabhan77,gron90,brevik94,brevik04,cataldo05,brevik06,brevik02,li09,brevik10a,sebastiani10,velten12,velten13,velten13a,bamba16,capozziello06a,nojiri07}
(for a review, see \cite{bre})
in which various aspects of the theory of viscous cosmology have been discussed. A
number of the articles  deal with multiple coupled viscous fluids
\cite{wang14,balakin11,nunes11}. Various cosmological scenarios of
the coupling between energy and matter were explored in the review
\cite{bolotin15}, and some examples of inhomogeneous viscous coupled fluids
were considered in \cite{bamba12,elizalde14,brevik15}. Bounce cosmology,
induced by an inhomogeneous viscous fluid, was analyzed in
\cite{brevik14}, and inflation as produced by two coupled fluids
in the presence of viscosity was considered in \cite{brevik16}, taking into
account the correspondence between parameters recently obtained in the analysis
of the Planck satellite data. Lately, conditions for the existence of an
inflationary universe with a bulk viscosity, and at the same time preventing
the occurrence of self-reproduction, were derived in \cite{brevik17}.

Cosmic fluids obeying a van der Waals equation of state were investigated in
\cite{capozziello02,kremer03,capozziello03,kremer04,khurshudyan14}. The van der
Waals fluid model can account both for the early and late-time accelerated
expansion stages of the universe. That is, inflation may be described by a van
der Waals fluid, with the specific properties of a cosmic fluid obeying the van der
Waals equation of state having been analyzed \cite{vardiasli17}. Various
versions of the van der Waals equation can be considered in seeking a better
match with the observational data of the Planck satellite \cite{jantsch16}.
However, generally speaking, it is not so easy to get a good consistency with the Planck data in
this case.

In the present work we will study the inflationary expansion of our early
universe in terms of the van der Waals equation, as equation of state for the
cosmic fluid, where a bulk viscosity contribution will be assumed to be
present. We will discuss a number of theoretical models along this line, with the
aim to obtain the most perfect agreement with the Planck data.  Although
inflation itself can actually be described in the absence of viscosity, it will
be our main goal to investigate the specific influence of viscous effects on
the resulting inflationary parameters. In particular, we will obtain some
examples where the agreement with Planck data can be achieved only in the
presence of viscosity.

\section{Van der Waals fluid for inflation in the presence of viscosity}

In this section we will investigate the inflationary universe by making use of
the formalism for an inhomogeneous viscous fluid, in a spatially flat
Friedmann-Lema\^{\i}tre-Robertson-Walker spacetime. We will describe inflation in terms of a
van der Waals equation of state, and  include a bulk viscosity term. This
is in fact a brand new element in our analysis: the inclusion of viscosity
in van der Waals models for the inflationary epoch.

The first Friedmann equation for the Hubble rate is
\begin{equation}
\rho=\frac{3}{k^2}H^2, \label{1}
\end{equation}
where $\rho$ is the energy density, $H(t)=\dot{a}(t)/a(t)$ the Hubble
parameter, $a(t)$ the scale factor, and $k^2=8\pi G$ with $G$ being Newton's
gravitational constant. A dot denotes derivative with respect to the cosmic time.

In a spatially flat Friedmann-Lema\^{\i}tre-Robertson-Walker universe, the metric is
\begin{equation}
ds^2=-dt^2+a^2(t)\sum_i dx_i^2. \label{2}
\end{equation}
We will describe the inflationary universe filled with a fluid satisfying
a nonlinear
inhomogeneous equation of state,
\begin{equation}
p=\omega(\rho,t)\rho+f(\rho)-3H\zeta(H.t), \label{3}
\end{equation}
where the thermodynamic parameter $\omega(\rho,t)$ depends on the energy
density and time, $\zeta(H,t)$ is the bulk viscosity, dependent on the Hubble
parameter and time, and $f(\rho)$ is, in the general case, an arbitrary
function. The case $f(\rho)=0$ was examined in \cite{brevik16}. Generally
speaking, an effective equation of state of this kind is quite typical in
modified gravity theories (see, e.g., \cite{nojiri07a,capo} for review).

We can write the energy conservation law as
\begin{equation}
\dot{\rho} +\frac{3\dot{a}}{a}(\rho+p)=0. \label{4}
\end{equation}
Note that this equation has the same form as the corresponding one for a
nonviscous fluid. Usually, in viscous cosmology, the equation would include a
term on the right hand side containing the bulk viscosity. But, we should
emphasize that the properties coming from viscosity are here included through the
inhomogeneous equation of state, instead of via a more standard bulk
viscosity term.

In the following, we will study the van der Waals fluid model
\cite{vardiasli17} in cases where the inhomogeneous viscous fluid theory is
applied right in the inflationary period. It is convenient to recall some
basic characteristics of the van der Waals theory in general. This description
can be viewed as an intermediate theory, which interpolates qualitatively the
transition from a gas to a liquid. For a dilute gas it must reduce to the
equation of state of an ideal gas, and as the density increases it must yield
the compressibility of a liquid. In other words, this particular equation of
state describes the behavior of the fluid in the intermediate region.
Physically, this means that the finite volume of the microscopic  constituents
(molecules in the case of a gas) is taken into account (for a more detailed
treatment, see Ref.~\cite{landau87}.)

In cosmology, there seems to be no apparent reason why the van der Waals
equation of state should be particularly preferred, as compared to other
alternatives.  However, equations of this sort have proved to be an effective
and very simple variant of a whole class of interpolating formulas in ordinary
hydrodynamics. Furthermore, a phase transition is expected to be related
with the very early universe, especially with its origin. This gives
ground to think of der Waals fluid models as a serious and reasonable
alternative in cosmological applications, too. Indeed, the corresponding
equation fits very naturally in, as a possibility for an inhomogeneous equation
of state.

\bigskip

Let us consider a cosmological fluid obeying a one-parameter van der Waals
equation of state, and include  also a bulk viscous term. We take the
thermodynamic parameter, the nonlinear function in the equation of state
(\ref{3}), and the bulk viscosity, to be of the form
\begin{align}
\omega(\rho,t)&= \frac{\gamma}{1-\beta \rho/\rho_c}, \nonumber \\
f(\rho)&=  -\frac{\alpha}{\rho_c}\rho^2,  \nonumber \\
\zeta(H,t)&=\xi_1(t)(3H)^n.  \label{5}
\end{align}
This model contains three independent, constant parameters, namely $\alpha, \,
\beta$ and $\gamma$. The parameter $\alpha$ is  related to the intermolecular
interaction, $\gamma$ describes the particle size, while $\beta$ is a critical
thermodynamic parameter.

The critical  value $\rho_c$ indicates that cosmic fluids change phases under
certain thermodynamic conditions. Whereas perfect fluids do not permit phase
transitions to occur, here the phase transition phenomenon can be taken care of
by means of the two-phase fluid described by the van der Waals equation.

Further, we consider the case $n=1$ for the bulk viscosity, and choose
$\xi_1(t)=\tau$, where $\tau$ is a positive constant. Then, the bulk viscosity
$\zeta(H,t)=3H\tau$ becomes a linear function of the Hubble parameter and
equation (\ref{3}) has the following form
\begin{equation}
p=\frac{\gamma \rho}{1-\beta \rho/\rho_c}-\frac{\alpha}{\rho_c}\rho^2-3k^2\tau
\rho. \label{6}
\end{equation}
Next, using equations (\ref{4}) and (\ref{6}) one obtains the following
differential equation relating the scale factor to the energy density:
\begin{equation}
a\left( 1-\beta \frac{\rho}{\rho_c}\right) \frac{d\rho}{da}+3\rho \left[ \gamma
-3k^2\tau+1+(3k^2\beta \tau-\alpha-\beta)\frac{\rho}{\rho_c}+\frac{\alpha
\beta}{\rho_c^2}\rho^2\right]=0. \label{7}
\end{equation}
Using the Friedmann equation (\ref{1}), we may rewrite this equation in the
form
\begin{equation}
\frac{da}{a}=-\frac{(1-\beta x)dx}{x(Ax^2+Bx+C)}. \label{8}
\end{equation}
We have here introduced the scaled variable $x=\rho/\rho_c$, and denoted
$A=\alpha \beta, \, B=3k^2\beta \tau-\alpha -\beta, \, C=\gamma-3k^2\tau +1$.
It is seen  that the right hand side of Eq.~(\ref{8}) contains the polynomial
$Ax^2+Bx+C$.

We will consider various cases for the integration of Eq.~(\ref{8}), depending
on the sign of the discriminant $4AC-B^2$.

\bigskip
{\it Case 1:} $4AC-B^2<0$.

\bigskip

The evolution of the scale factor as a function of the energy density becomes
\begin{equation}
a=a_0\left[ \frac{1}{x}\left(
x^2+\frac{B}{A}x+\frac{C}{A}\right)\right]^{1/2C}\exp \left(\frac{2\beta
+\frac{B}{C}}{\sqrt{4AC-B^2}}{\rm{arctan}} \frac{2Ax+B}{\sqrt{4AC-B^2}}\right),
\label{9}
\end{equation}
where $a_0$ is an integration constant.

If we put, for simplicity,  $\beta =-(B/2C)$,  we obtain
\begin{equation}
a^C=a_0\frac{\left( x^2+\frac{B}{A}+\frac{C}{A}\right)^{1/2}}{x}. \label{10}
\end{equation}
If $A=C=1/2$, we can calculate all parameters of this model,
\begin{align}
\gamma &=3\tau k^2-\frac{1}{2}, \nonumber \\
\alpha &=\pm \sqrt{\frac{3}{2}\tau k^2-1}, \nonumber \\
\beta= &\pm \frac{1}{\sqrt{2(3\tau k^2-2)}}. \label{11}
\end{align}

In this particular case the scale factor becomes
\begin{equation}
a=a_0\frac{x^2-2\beta x+1}{x^2}, \label{12}
\end{equation}
and we can write the corresponding van der Waals equation of state as
\begin{equation}
p=\frac{\left( 3\tau k^2-\frac{1}{2}\right) \rho}{1-\frac{x}{\sqrt{2(3\tau
k^2-2)}}}
-  \frac{\sqrt{\frac{3}{2}\tau k^2-1}}{\rho_c}\rho^2 - 3k^2\tau \rho.
\end{equation}

\bigskip

{\it Case 2:} $4AC-B^2=0.$

\bigskip

This is a simpler case, for which the solution of Eq.~(\ref{8}) is
\begin{equation}
a=a_0\left( \frac{x+\frac{B}{2A}}{x}\right)^{4A^2/B^2}\exp \left(
-\frac{\frac{2}{B}+\frac{\beta}{A}}{x+\frac{B}{2A}}\right). \label{14}
\end{equation}
If we impose $\frac{2}{B}+\frac{\beta}{A}=0$,  then $B=-2\alpha$ and the ratio
of the parameters becomes $\alpha/\beta=1-3k^2\tau$. The expression for the
scale factor  becomes then
\begin{equation}
a=a_0\left( \frac{x-1/\beta}{x}\right)^{\beta/\alpha}. \label{15}
\end{equation}
In the early universe, when $a\rightarrow 0$, the initial value of the energy
density reads $\rho_{\rm in}=\beta^{-1}\rho_c$.

\bigskip
{\it Case 3:} $4AC-B^2>0$

\bigskip
This case leads to the result
\begin{equation}
a=\frac{a_0}{x^{1/C}}\left[ \frac{|x-x_2|^{\beta-1/x_2}}{|x-x_1|^{\beta-1/x_1}}
\right]^{\frac{1}{\sqrt{B^2-4AC}}}, \label{16}
\end{equation}
where $ x_1=-\frac{B}{2A} -\tilde{\alpha}, \quad
x_2=-\frac{B}{2A}+\tilde{\alpha}$, and $
\tilde{\alpha}=\frac{\sqrt{B^2-4AC}}{2A}.$
The solution (\ref{16}) contains a singularity at $x=x_1$.

It is of interest to consider the following particular case of this model. We
choose the values $\alpha=9\gamma/8$ and $\beta=1/3$ for the parameters in the
van der Waals equation of state, and choose now another value $n=2$ in the
expression (\ref{5}) for the bulk viscosity. As before, we take
$\xi_1(t)=\tau$, with $\tau$ a positive constant.

Then the bulk viscosity $\zeta(H,t)=3\tau H^2$ becomes a quadratic function of
the Hubble parameter, and the equation of state (\ref{3}) takes the form
\begin{equation}
p=\frac{\gamma
\rho}{1-\frac{\rho}{3\rho_c}}-\frac{9\gamma}{8\rho_c}\rho^2-3k^2\tau \rho.
\label{17}
\end{equation}
Next, using Eqs.~(\ref{4}) and (\ref{16}), we obtain the following differential
equation for the scale factor and energy density,
\begin{equation}
a\left( 1-\frac{1}{3}\frac{\rho}{\rho_c}\right) \frac{d\rho}{da}+3\rho \left[
\gamma-3k^2\tau +1+\frac{1}{3\rho_c}\left(
3k^2\tau-\frac{27}{8}\gamma-1\right)\rho+\frac{3}{8}\frac{\gamma}{\rho_c^2}\rho^2\right]=0.
\label{18}
\end{equation}
This equation can be transformed as
\begin{equation}
\frac{da}{a}=-\frac{\left( 1-\frac{1}{3}x\right) dx}{x(Ax^2+Bx+C)}, \label{19}
\end{equation}
with the scaled variable $x=\rho/\rho_c$. Moreover $A=3\gamma/8, \,
B=k^2\tau-9\gamma/8-1/3, \, C=\gamma-3k^2\tau +1.$

In analogy to our treatment above, we consider various cases for the
integration of Eq.~(\ref{19}).

\bigskip
{\it Case 1:
$4AC-B^2<0.$}

\bigskip

The scale factor as a function of $x$ becomes
\begin{equation}
a=a_0\left[\frac{1}{x}\left(
x^2+\frac{B}{A}x+\frac{C}{A}\right)^{1/2}\right]^{1/C}\exp
\left(\frac{\frac{2}{3}-\frac{B}{C}}{\sqrt{4AC-B^2}}{\rm
arctan}\frac{2Ax+B}{\sqrt{4AC-B^2}}\right), \label{20}
\end{equation}
where $a_0$ is an integration constant. If we suppose that $B/C=2/3$, then
$\gamma=\frac{8}{9}\left(\frac{1}{3}-k^2\tau\right)$, and the solution
(\ref{20}) takes the simple form
\begin{equation}
a^C=\frac{a_0}{x}\left( x^2+\frac{B}{A}x+\frac{C}{A}\right)^{1/2}. \label{21}
\end{equation}

\bigskip
{\it Case 2:
$4AC-B^2=0.$}

\bigskip

In this case, the scale factor is
\begin{equation}
a=a_0\left( \frac{1}{x}\big|x+\frac{B}{2A}\big| \right)^{4A^2/B^2}\exp \left(
-\frac{\frac{1}{3}+\frac{2A}{B}}{\big|x+\frac{B}{2A}\big|}\right). \label{22}
\end{equation}
In the particular case $\frac{1}{3}+\frac{2A}{B}=0$, one obtains
\begin{equation}
a=a_0\left( \frac{|x-3|}{x}\right)^{1/9}. \label{23}
\end{equation}
We see that, at the beginning of inflation, $a\rightarrow 0$, the initial value
of the energy density is $\rho_{\rm in}=3\rho_c$.

\bigskip
{\it Case 3:
$4AC-B^2 >0.$}

\bigskip

The scale factor becomes now
\begin{equation}
a^A = a_0\, x^{\frac{C}{A}}\left(
\frac{|x-x_1|^{1-3/x_1}}{|x-x_2|^{1-3/x_2}}\right)^{\frac{1}{3(x_2-x_1)}},
\label{24}
\end{equation}
where $x_{1,2}=-\frac{B}{2A} \pm \alpha, \, \alpha=\frac{\sqrt{B^2-4AC}}{2A}$.
This expression contains a singularity at $x=x_2$.
Hence, we have derived an analytic expression for the scale factor in terms
of the energy density for different choices of the parameters of the theory.

\section{Comparison with Planck observations}

We will now compare the predictions of our inflationary model with the latest
Planck satellite observational data. In particular, we will calculate the
parameters of inflation, and consider how the spectral index and the
tensor-to-scalar ratio match with the values obtained from the astronomical
data analysis.

For simplicity we consider the evolution of the scale factor for the viscous
van der Waals fluid, when $n=1$ and the discriminant satisfies the condition
$4AC-B^2=0$. The scale factor is then given by the expression (\ref{15}). We
intend to see under what conditions the inflationary model complies with the
Planck data.

First, let us calculate the "slow-roll" parameters $\varepsilon$ and $\eta$,
\begin{equation}
\varepsilon=-\frac{\dot{H}}{H^2}=-\frac{\alpha}{2}\left( x-\frac{1}{\beta}\right), \quad
\eta= \varepsilon - \frac{1}{2 \varepsilon H} \dot{\varepsilon} =-\frac{\alpha}{2}\left( 2x-\frac{1}{\beta}\right). \label{25}
\end{equation}
Note that, at the beginning of inflation, we have
$\varepsilon(\beta^{-1}\rho_c)=0$ and
$|\eta(\beta^{-1}\rho_c)|=\frac{\alpha}{2\beta}$. The requirement
$|\eta(\beta^{-1}\rho_c)| \ll 1$ leads to the inequality $\frac{\alpha}{\beta}
\ll 2$. In this way, the "slow-roll" conditions can be satisfied.

From the "slow-roll" parameters one gets for the spectral index $n_s$,
\begin{equation}
n_s-1= -6\varepsilon +2\eta =\alpha \left( x-\frac{2}{\beta}\right). \label{26}
\end{equation}
From the Planck data one has for the tensor-to-scalar ratio $r=16\varepsilon
\leq 0.11$. One then obtains $x \geq -\frac{1}{\alpha}\,0.01375
+\frac{1}{\beta}$.  For the allowed region, we thus have $n_s-1 \approx -0.0(3)
\geq -0.01375-\frac{\alpha}{\beta}$. Consequently, in order to comply with the
observations, one needs to require $\frac{\alpha}{\beta} \geq 0.01958.$
For this case without viscosity ($\tau =0$) we get $\alpha =\beta$  and $\left| \eta (\alpha^{-1} \rho_c) \right|=\frac{1}{2}$, what violates the basic condition $\eta \ll 1$. Thus, viscosity proves to be absolutely necessary in this situation.

Now, we consider the scale factor (\ref{22}) in the particular case discussed
above of the van der Waals model. A calculation of the "slow-roll" parameters
for the corresponding inflation yields
\begin{equation}
\varepsilon =\frac{3\gamma}{4}\frac{|x-3|}{|x-3|-x},          \quad
\eta=\frac{3\gamma}{4}\frac{|x-3|+x}{|x-3|-x}. \label{27}
\end{equation}
The initial values of these parameters are $\varepsilon(3\rho_c)=0$ and
$|\eta(3\rho_c)|=3\gamma/4$. The first condition $\varepsilon \ll 1$, relevant
at the beginning of inflation, is fulfilled, and the second  condition $\eta
\ll 1$ is valid provided $\gamma \ll 4/3$.

The spectral index is given by the expression
\begin{equation}
n_s-1=     -3\gamma \left( 1+ \frac{1}{2}\frac{x}{|x-3|-x}\right).
\label{28}
\end{equation}
Since $ r\leq 0.11$, one must have $\frac{x}{|x-3|-x} \leq 0.0092\gamma^{-1} -1$ and
$n_s-1 \approx -0.0(3) \leq - \frac{3}{2}\gamma (1+0.0092\gamma^{-1} ) $, or
equivalently $0.0(1) \geq \frac{1}{2}\gamma + 0.0046$.  Thus, in order to
comply with the results of Planck, we must require $\gamma \leq 0.013$.
For this particular case without viscosity we get $\gamma = \frac{8}{9}$  and $\left| \eta (3 \rho_c) \right|=\frac{2}{3}$ . The restriction $\gamma \leq 0.013$ violates the basic conditions to be fulfilled and, once more, this case in the absence of viscosity does not match the Planck data, rendering viscosity necessary.

As a consequence, under some simple restrictions, our model is perfectly able to
comply with the astronomical data. We conclude that the modified van der Waals
equation of state, as postulated here, can adequately describe the
inflationary universe, in a natural and standard way.

\section{Conclusions}

We have investigated in this paper theoretical fluid models for inflation in the
presence of viscosity, when the fluid obeys the van der Waals equation of
state. In cosmology, there seems to be no apparent reason why the van der Waals
equation of state should be particularly preferred, as compared to other
alternatives.  However, equations of this sort have proved to be an effective
and very simple variant of a whole class of interpolating formulas in ordinary
hydrodynamics an this gives ground to think of them as a serious and reasonable
alternative in cosmological applications. In the paper, a flat
Friedmann-Lema\^{\i}tre-Robertson-Walker spacetime has been assumed. It is worth noticing
that the van der Waals equation is in fact a particular case of an effective
equation of state in modified gravity. We have parameterized the van der Waals
model and found solutions of the gravitational equations of motion, in the form
of a scale factor as a function of the energy density.

The validity of this model has been tested via comparison of the most
fundamental inflationary parameters, such as the spectral index and the
tensor-to-scalar ratio, calculated within our model, with the latest results
obtained from the Planck satellite data analysis. It has been shown above that,
after complying with some restrictions on the parameters in the equation of
state, one obtains indeed good agreement with the astronomical observations.

Usually, when considering the inflationary epoch the effect of viscosity is
neglected, because the bulk viscosity is regarded to be small. However, it is
clear from our results that the inclusion of viscosity in the van der Waals
fluid model yields an increased flexibility in the mathematical formalism and
also facilitates the task of getting  agreement with the  Planck results, in a
smooth natural way.

When trying to predict the future development of our universe, within the
framework of viscous cosmology, it is of crucial interest to know the
approximate value of the bulk viscosity, $\zeta_0$, at present. In this
context, we should mention that the analysis of various experiments carried out
in Refs.~\cite{normann16} leads to the following preferred value
\begin{equation}
\zeta_0  \approx 10^6~{\rm Pa~s}. \label{29}
\end{equation}

\medskip

\noindent {\bf Acknowledgements}.  EE and SDO are partially supported by CSIC,
I-LINK1019 Project, by MINECO (Spain), Projects FIS2013-44881-P and
FIS2016-76363-P,  and by the CPAN Consolider Ingenio Project. This work is
partially supported by the Russian Ministry of Education and Science, project No. 3.1386.2017. The paper was finished while EE was visiting the Kobayashi-Maskawa Institute in Nagoya, Japan. EE is much obligued with S. Nojiri, T. Katsuragawa, and the rest of the members of the KMI for the very warm hospitality.

\end{document}